\documentclass[conference]{IEEEtran}

\usepackage{amsmath,amsfonts,mathtools}
\usepackage{pifont}
\usepackage{algorithmic}
\usepackage{algorithm}
\usepackage{array}
\usepackage{textcomp}
\usepackage{stfloats}
\usepackage{url}
\usepackage{verbatim}
\usepackage{graphicx}
\usepackage{gensymb}
\usepackage{cite}
\usepackage{cases}
\usepackage{multirow}
\usepackage{makecell}
\usepackage{color, colortbl}
\definecolor{mixed}{gray}{0.9}
\definecolor{synth}{gray}{0.8}
\usepackage{multicol}
\usepackage[tableposition=top]{caption}
\usepackage{subcaption}
\usepackage{siunitx}

\usepackage{colortbl}
\usepackage{pgfplots}
\usepackage{pgfplotstable}
\usepgfplotslibrary{fillbetween}
\usepgfplotslibrary{external}
\usepackage{longtable}
\usepackage{xspace}
\usepackage{arydshln}

\usepackage{svg}
\usepackage{booktabs}

\usepackage{soul}

\usepackage[font=scriptsize]{subcaption}
\usepackage[font=footnotesize]{caption}

\usepackage[outdir=./]{epstopdf}
\usepackage{tikz}
\pgfplotsset{compat=newest}
\pgfplotsset{plot coordinates/math parser=false}
\newlength\fheight
\newlength\fwidth
\usetikzlibrary{plotmarks,shapes,patterns,decorations.pathreplacing,backgrounds,calc,arrows,arrows.meta,spy,matrix,external}
\usepackage{tikzscale}
\usepackage[utf8]{inputenc}

\pgfplotsset{
    compat=1.15,
    my axis style/.style={
        ylabel style ={font=\footnotesize},
        xlabel style ={font=\footnotesize},
        xmajorgrids,
        ymajorgrids,
        legend style={legend cell align=left, align=left, draw=white!15!black,font=\footnotesize, anchor=north west, at={(0.01,1.02)}},
        legend columns=3
    },
}

\pgfplotsset{every tick label/.append style={font=\scriptsize}}

\usepackage[capitalise]{cleveref}
\crefname{section}{Sec.}{Secs.}
\crefname{figure}{Fig.}{Figs.}

\usepackage{eqparbox}

\DeclareCaptionLabelFormat{andtable}{#1~#2  \&  \tablename~\thetable}

\tikzset{
    export as png/.style={
        external/system call/.add={}{
            && convert -density #1 -transparent white "\image.pdf" "\image.png"
        },
    },
    export as png/.default={200},
}
\usepackage[acronyms,nonumberlist,nopostdot,nomain,nogroupskip,acronymlists={hidden}]{glossaries}
\newglossary[algh]{hidden}{acrh}{acnh}{Hidden Acronyms}

\newacronym{3gpp}{3GPP}{3rd Generation Partnership Project}
\newacronym{4g}{4G}{4th generation}
\newacronym{5g}{5G}{5th generation}
\newacronym{6g}{6G}{6th generation}
\newacronym{5gc}{5GC}{5G Core}
\newacronym{adc}{ADC}{Analog to Digital Converter}
\newacronym{aerpaw}{AERPAW}{Aerial Experimentation and Research Platform for Advanced Wireless}
\newacronym{ai}{AI}{Artificial Intelligence}
\newacronym{aimd}{AIMD}{Additive Increase Multiplicative Decrease}
\newacronym{am}{AM}{Acknowledged Mode}
\newacronym{amc}{AMC}{Adaptive Modulation and Coding}
\newacronym{amf}{AMF}{Access and Mobility Management Function}
\newacronym{aops}{AOPS}{Adaptive Order Prediction Scheduling}
\newacronym{api}{API}{Application Programming Interface}
\newacronym{apn}{APN}{Access Point Name}
\newacronym{ap}{AP}{Application Protocol}
\newacronym{aqm}{AQM}{Active Queue Management}
\newacronym{ar}{AR}{Augmented Reality}
\newacronym{ausf}{AUSF}{Authentication Server Function}
\newacronym{avc}{AVC}{Advanced Video Coding}
\newacronym{awgn}{AGWN}{Additive White Gaussian Noise}
\newacronym{balia}{BALIA}{Balanced Link Adaptation Algorithm}
\newacronym{bbu}{BBU}{Base Band Unit}
\newacronym{bdp}{BDP}{Bandwidth-Delay Product}
\newacronym{ber}{BER}{Bit Error Rate}
\newacronym{bf}{BF}{Beamforming}
\newacronym{bler}{BLER}{Block Error Rate}
\newacronym{brr}{BRR}{Bayesian Ridge Regressor}
\newacronym{bs}{BS}{Base Station}
\newacronym{bsr}{BSR}{Buffer Status Report}
\newacronym{bss}{BSS}{Business Support System}
\newacronym{ca}{CA}{Carrier Aggregation}
\newacronym{caas}{CaaS}{Connectivity-as-a-Service}
\newacronym{cb}{CB}{Code Block}
\newacronym{cc}{CC}{Congestion Control}
\newacronym{ccid}{CCID}{Congestion Control ID}
\newacronym{cco}{CC}{Carrier Component}
\newacronym{cdd}{CDD}{Cyclic Delay Diversity}
\newacronym{cdf}{CDF}{Cumulative Distribution Function}
\newacronym{cdn}{CDN}{Content Distribution Network}
\newacronym{cn}{CN}{Core Network}
\newacronym{codel}{CoDel}{Controlled Delay Management}
\newacronym{comac}{COMAC}{Converged Multi-Access and Core}
\newacronym{cord}{CORD}{Central Office Re-architected as a Datacenter}
\newacronym{cornet}{CORNET}{COgnitive Radio NETwork}
\newacronym{cosmos}{COSMOS}{Cloud Enhanced Open Software Defined Mobile Wireless Testbed for City-Scale Deployment}
\newacronym{cots}{COTS}{Commercial Off-the-Shelf}
\newacronym{cp}{CP}{Control Plane}
\newacronym{cyp}{CP}{Cyclic Prefix}
\newacronym{up}{UP}{User Plane}
\newacronym{cpu}{CPU}{Central Processing Unit}
\newacronym{cqi}{CQI}{Channel Quality Information}
\newacronym{cr}{CR}{Cognitive Radio}
\newacronym{cran}{C-RAN}{Cloud \gls{ran}}
\newacronym{crs}{CRS}{Cell Reference Signal}
\newacronym{csi}{CSI}{Channel State Information}
\newacronym{csirs}{CSI-RS}{Channel State Information - Reference Signal}
\newacronym{cu}{CU}{Central Unit}
\newacronym{d2tcp}{D$^2$TCP}{Deadline-aware Data center TCP}
\newacronym{d3}{D$^3$}{Deadline-Driven Delivery}
\newacronym{dac}{DAC}{Digital to Analog Converter}
\newacronym{dag}{DAG}{Directed Acyclic Graph}
\newacronym{das}{DAS}{Distributed Antenna System}
\newacronym{dash}{DASH}{Dynamic Adaptive Streaming over HTTP}
\newacronym{dc}{DC}{Dual Connectivity}
\newacronym{dccp}{DCCP}{Datagram Congestion Control Protocol}
\newacronym{dce}{DCE}{Direct Code Execution}
\newacronym{dci}{DCI}{Downlink Control Information}
\newacronym{dctcp}{DCTCP}{Data Center TCP}
\newacronym{dl}{DL}{Downlink}
\newacronym{dmr}{DMR}{Deadline Miss Ratio}
\newacronym{dmrs}{DMRS}{DeModulation Reference Signal}
\newacronym{drlcc}{DRL-CC}{Deep Reinforcement Learning Congestion Control}
\newacronym{drs}{DRS}{Discovery Reference Signal}
\newacronym{du}{DU}{Distributed Unit}
\newacronym{e2e}{E2E}{end-to-end}
\newacronym{ecaas}{ECaaS}{Edge-Cloud-as-a-Service}
\newacronym{ecn}{ECN}{Explicit Congestion Notification}
\newacronym{edf}{EDF}{Earliest Deadline First}
\newacronym{embb}{eMBB}{Enhanced Mobile Broadband}
\newacronym{empower}{EMPOWER}{EMpowering transatlantic PlatfOrms for advanced WirEless Research}
\newacronym{enb}{eNB}{evolved Node Base}
\newacronym{endc}{EN-DC}{E-UTRAN-\gls{nr} \gls{dc}}
\newacronym{epc}{EPC}{Evolved Packet Core}
\newacronym{eps}{EPS}{Evolved Packet System}
\newacronym{es}{ES}{Edge Server}
\newacronym{etsi}{ETSI}{European Telecommunications Standards Institute}
\newacronym[firstplural=Estimated Times of Arrival (ETAs)]{eta}{ETA}{Estimated Time of Arrival}
\newacronym{eutran}{E-UTRAN}{Evolved Universal Terrestrial Access Network}
\newacronym{faas}{FaaS}{Function-as-a-Service}
\newacronym{fapi}{FAPI}{Functional Application Platform Interface}
\newacronym{fdd}{FDD}{Frequency Division Duplexing}
\newacronym{fdm}{FDM}{Frequency Division Multiplexing}
\newacronym{fdma}{FDMA}{Frequency Division Multiple Access}
\newacronym{fed4fire}{FED4FIRE+}{Federation 4 Future Internet Research and Experimentation Plus}
\newacronym{fir}{FIR}{Finite Impulse Response}
\newacronym{fit}{FIT}{Future \acrlong{iot}}
\newacronym{fpga}{FPGA}{Field Programmable Gate Array}
\newacronym{fr2}{FR2}{Frequency Range 2}
\newacronym{fs}{FS}{Fast Switching}
\newacronym{fscc}{FSCC}{Flow Sharing Congestion Control}
\newacronym{ftp}{FTP}{File Transfer Protocol}
\newacronym{fw}{FW}{Flow Window}
\newacronym{ge}{GE}{Gaussian Elimination}
\newacronym{gnb}{gNB}{Next Generation Node Base}
\newacronym{gop}{GOP}{Group of Pictures}
\newacronym{gpr}{GPR}{Gaussian Process Regressor}
\newacronym{gpu}{GPU}{Graphics Processing Unit}
\newacronym{gtp}{GTP}{GPRS Tunneling Protocol}
\newacronym{gtpc}{GTP-C}{GPRS Tunnelling Protocol Control Plane}
\newacronym{gtpu}{GTP-U}{GPRS Tunnelling Protocol User Plane}
\newacronym{gtpv2c}{GTPv2-C}{\gls{gtp} v2 - Control}
\newacronym{gw}{GW}{Gateway}
\newacronym{harq}{HARQ}{Hybrid Automatic Repeat reQuest}
\newacronym{hetnet}{HetNet}{Heterogeneous Network}
\newacronym{hh}{HH}{Hard Handover}
\newacronym{hol}{HOL}{Head-of-Line}
\newacronym{hqf}{HQF}{Highest-quality-first}
\newacronym{hss}{HSS}{Home Subscription Server}
\newacronym{http}{HTTP}{HyperText Transfer Protocol}
\newacronym{ia}{IA}{Initial Access}
\newacronym{iab}{IAB}{Integrated Access and Backhaul}
\newacronym{ic}{IC}{Incident Command}
\newacronym{ietf}{IETF}{Internet Engineering Task Force}
\newacronym{imsi}{IMSI}{International Mobile Subscriber Identity}
\newacronym{imt}{IMT}{International Mobile Telecommunication}
\newacronym{iot}{IoT}{Internet of Things}
\newacronym{ip}{IP}{Internet Protocol}
\newacronym{itu}{ITU}{International Telecommunication Union}
\newacronym{kpi}{KPI}{Key Performance Indicator}
\newacronym{kpm}{KPM}{Key Performance Measurement}
\newacronym{kvm}{KVM}{Kernel-based Virtual Machine}
\newacronym{los}{LOS}{Line-of-Sight}
\newacronym{lsm}{LSM}{Link-to-System Mapping}
\newacronym{lstm}{LSTM}{Long Short Term Memory}
\newacronym{lte}{LTE}{Long Term Evolution}
\newacronym{lxc}{LXC}{Linux Container}
\newacronym{m2m}{M2M}{Machine to Machine}
\newacronym{mac}{MAC}{Medium Access Control}
\newacronym{manet}{MANET}{Mobile Ad Hoc Network}
\newacronym{mano}{MANO}{Management and Orchestration}
\newacronym{mc}{MC}{Multi-Connectivity}
\newacronym{mcc}{MCC}{Mobile Cloud Computing}
\newacronym{mchem}{MCHEM}{Massive Channel Emulator}
\newacronym{mcs}{MCS}{Modulation and Coding Scheme}
\newacronym{mec}{MEC}{Multi-access Edge Computing}
\newacronym{mec2}{MEC}{Mobile Edge Cloud}
\newacronym{mfc}{MFC}{Mobile Fog Computing}
\newacronym{mgen}{MGEN}{Multi-Generator}
\newacronym{mi}{MI}{Mutual Information}
\newacronym{mib}{MIB}{Master Information Block}
\newacronym{miesm}{MIESM}{Mutual Information Based Effective SINR}
\newacronym{mimo}{MIMO}{Multiple Input, Multiple Output}
\newacronym{ml}{ML}{Machine Learning}
\newacronym{mlr}{MLR}{Maximum-local-rate}
\newacronym[plural=\gls{mme}s,firstplural=Mobility Management Entities (MMEs)]{mme}{MME}{Mobility Management Entity}
\newacronym{mmtc}{mMTC}{Massive Machine-Type Communications}
\newacronym{mmwave}{mmWave}{millimeter wave}
\newacronym{mpdccp}{MP-DCCP}{Multipath Datagram Congestion Control Protocol}
\newacronym{mptcp}{MPTCP}{Multipath TCP}
\newacronym{mr}{MR}{Maximum Rate}
\newacronym{mrdc}{MR-DC}{Multi \gls{rat} \gls{dc}}
\newacronym{mse}{MSE}{Mean Square Error}
\newacronym{mss}{MSS}{Maximum Segment Size}
\newacronym{mt}{MT}{Mobile Termination}
\newacronym{mtd}{MTD}{Machine-Type Device}
\newacronym{mtu}{MTU}{Maximum Transmission Unit}
\newacronym{mumimo}{MU-MIMO}{Multi-user \gls{mimo}}
\newacronym{mvno}{MVNO}{Mobile Virtual Network Operator}
\newacronym{nalu}{NALU}{Network Abstraction Layer Unit}
\newacronym{nas}{NAS}{Non-Access Stratum}
\newacronym{nbiot}{NB-IoT}{Narrow Band IoT}
\newacronym{nfv}{NFV}{Network Function Virtualization}
\newacronym{nfvi}{NFVI}{Network Function Virtualization Infrastructure}
\newacronym{ngrg}{nGRG}{next Generation Research Group}
\newacronym{ni}{NI}{Network Interfaces}
\newacronym{nic}{NIC}{Network Interface Card}
\newacronym{nlos}{NLOS}{Non-Line-of-Sight}
\newacronym{now}{NOW}{Non Overlapping Window}
\newacronym{nsm}{NSM}{Network Service Mesh}
\newacronym{nr}{NR}{New Radio}
\newacronym{nrf}{NRF}{Network Repository Function}
\newacronym{nsa}{NSA}{Non Stand Alone}
\newacronym{nse}{NSE}{Network Slicing Engine}
\newacronym{nssf}{NSSF}{Network Slice Selection Function}
\newacronym{o2i}{O2I}{Outdoor to Indoor}
\newacronym{oai}{OAI}{OpenAirInterface}
\newacronym{oaicn}{OAI-CN}{\gls{oai} \acrlong{cn}}
\newacronym{oairan}{OAI-RAN}{\acrlong{oai} \acrlong{ran}}
\newacronym{oam}{OAM}{Operations, Administration and Maintenance}
\newacronym{ofdm}{OFDM}{Orthogonal Frequency Division Multiplexing}
\newacronym{olia}{OLIA}{Opportunistic Linked Increase Algorithm}
\newacronym{omec}{OMEC}{Open Mobile Evolved Core}
\newacronym{onap}{ONAP}{Open Network Automation Platform}
\newacronym{onf}{ONF}{Open Networking Foundation}
\newacronym{onos}{ONOS}{Open Networking Operating System}
\newacronym{oom}{OOM}{\gls{onap} Operations Manager}
\newacronym{opnfv}{OPNFV}{Open Platform for \gls{nfv}}
\newacronym{oran}{O-RAN}{Open \gls{ran}}
\newacronym{orbit}{ORBIT}{Open-Access Research Testbed for Next-Generation Wireless Networks}
\newacronym{os}{OS}{Operating System}
\newacronym{oss}{OSS}{Operations Support System}
\newacronym{otic}{OTIC}{Open Testing \& Integration Centre}
\newacronym{pa}{PA}{Position-aware}
\newacronym{pase}{PASE}{Prioritization, Arbitration, and Self-adjusting Endpoints}
\newacronym{pawr}{PAWR}{Platforms for Advanced Wireless Research}
\newacronym{pbch}{PBCH}{Physical Broadcast Channel}
\newacronym{pcef}{PCEF}{Policy and Charging Enforcement Function}
\newacronym{pcfich}{PCFICH}{Physical Control Format Indicator Channel}
\newacronym{pcrf}{PCRF}{Policy and Charging Rules Function}
\newacronym{pdcch}{PDCCH}{Physical Downlink Control Channel}
\newacronym{pdcp}{PDCP}{Packet Data Convergence Protocol}
\newacronym{pdsch}{PDSCH}{Physical Downlink Shared Channel}
\newacronym{pdu}{PDU}{Packet Data Unit}
\newacronym{pf}{PF}{Proportional Fair}
\newacronym{pgw}{PGW}{Packet Gateway}
\newacronym{phich}{PHICH}{Physical Hybrid ARQ Indicator Channel}
\newacronym{phy}{PHY}{Physical}
\newacronym{pmch}{PMCH}{Physical Multicast Channel}
\newacronym{pmi}{PMI}{Precoding Matrix Indicators}
\newacronym{powder}{POWDER}{Platform for Open Wireless Data-driven Experimental Research}
\newacronym{ppo}{PPO}{Proximal Policy Optimization}
\newacronym{ppp}{PPP}{Poisson Point Process}
\newacronym{prach}{PRACH}{Physical Random Access Channel}
\newacronym{prb}{PRB}{Physical Resource Block}
\newacronym{psnr}{PSNR}{Peak Signal to Noise Ratio}
\newacronym{pss}{PSS}{Primary Synchronization Signal}
\newacronym{pucch}{PUCCH}{Physical Uplink Control Channel}
\newacronym{pusch}{PUSCH}{Physical Uplink Shared Channel}
\newacronym{qam}{QAM}{Quadrature Amplitude Modulation}
\newacronym{qci}{QCI}{\gls{qos} Class Identifier}
\newacronym{qoe}{QoE}{Quality of Experience}
\newacronym{qos}{QoS}{Quality of Service}
\newacronym{quic}{QUIC}{Quick UDP Internet Connections}
\newacronym{rach}{RACH}{Random Access Channel}
\newacronym{ran}{RAN}{Radio Access Network}
\newacronym[firstplural=Radio Access Technologies (RATs)]{rat}{RAT}{Radio Access Technology}
\newacronym{rcn}{RCN}{Research Coordination Network}
\newacronym{rc}{RC}{RAN Control}
\newacronym{rec}{REC}{Radio Edge Cloud}
\newacronym{red}{RED}{Random Early Detection}
\newacronym{renew}{RENEW}{Reconfigurable Eco-system for Next-generation End-to-end Wireless}
\newacronym{rf}{RF}{Radio Frequency}
\newacronym{rfc}{RFC}{Request for Comments}
\newacronym{rfr}{RFR}{Random Forest Regressor}
\newacronym{ric}{RIC}{\gls{ran} Intelligent Controller}
\newacronym{rlc}{RLC}{Radio Link Control}
\newacronym{rlf}{RLF}{Radio Link Failure}
\newacronym{rlnc}{RLNC}{Random Linear Network Coding}
\newacronym{rmr}{RMR}{RIC Message Router}
\newacronym{rmse}{RMSE}{Root Mean Squared Error}
\newacronym{rnis}{RNIS}{Radio Network Information Service}
\newacronym{rr}{RR}{Round Robin}
\newacronym{rrc}{RRC}{Radio Resource Control}
\newacronym{rrm}{RRM}{Radio Resource Management}
\newacronym{rru}{RRU}{Remote Radio Unit}
\newacronym{rs}{RS}{Remote Server}
\newacronym{rsrp}{RSRP}{Reference Signal Received Power}
\newacronym{rsrq}{RSRQ}{Reference Signal Received Quality}
\newacronym{rss}{RSS}{Received Signal Strength}
\newacronym{rssi}{RSSI}{Received Signal Strength Indicator}
\newacronym{rtt}{RTT}{Round Trip Time}
\newacronym{ru}{RU}{Radio Unit}
\newacronym{rw}{RW}{Receive Window}
\newacronym{rx}{RX}{Receiver}
\newacronym{s1ap}{S1AP}{S1 Application Protocol}
\newacronym{sa}{SA}{standalone}
\newacronym{sack}{SACK}{Selective Acknowledgment}
\newacronym{sap}{SAP}{Service Access Point}
\newacronym{sc2}{SC2}{Spectrum Collaboration Challenge}
\newacronym{scef}{SCEF}{Service Capability Exposure Function}
\newacronym{sch}{SCH}{Secondary Cell Handover}
\newacronym{scoot}{SCOOT}{Split Cycle Offset Optimization Technique}
\newacronym{sctp}{SCTP}{Stream Control Transmission Protocol}
\newacronym{sdap}{SDAP}{Service Data Adaptation Protocol}
\newacronym{sdk}{SDK}{Software Development Kit}
\newacronym{sdm}{SDM}{Space Division Multiplexing}
\newacronym{sdma}{SDMA}{Spatial Division Multiple Access}
\newacronym{sdn}{SDN}{Software-defined Networking}
\newacronym{sdr}{SDR}{Software-defined Radio}
\newacronym{seba}{SEBA}{SDN-Enabled Broadband Access}
\newacronym{sgsn}{SGSN}{Serving GPRS Support Node}
\newacronym{sgw}{SGW}{Service Gateway}
\newacronym{si}{SI}{Study Item}
\newacronym{sib}{SIB}{Secondary Information Block}
\newacronym{sinr}{SINR}{Signal to Interference plus Noise Ratio}
\newacronym{sip}{SIP}{Session Initiation Protocol}
\newacronym{siso}{SISO}{Single Input, Single Output}
\newacronym{sla}{SLA}{Service Level Agreement}
\newacronym{sm}{SM}{Service Model}
\newacronym{smf}{SMF}{Session Management Function}
\newacronym{smo}{SMO}{Service Management and Orchestration}
\newacronym{sms}{SMS}{Short Message Service}
\newacronym{smsgmsc}{SMS-GMSC}{\gls{sms}-Gateway}
\newacronym{snr}{SNR}{Signal-to-Noise-Ratio}
\newacronym{son}{SON}{Self-Organizing Network}
\newacronym{sptcp}{SPTCP}{Single Path TCP}
\newacronym{srb}{SRB}{Service Radio Bearer}
\newacronym{srn}{SRN}{Standard Radio Node}
\newacronym{srs}{SRS}{Sounding Reference Signal}
\newacronym{ss}{SS}{Synchronization Signal}
\newacronym{sss}{SSS}{Secondary Synchronization Signal}
\newacronym{st}{ST}{Spanning Tree}
\newacronym{svc}{SVC}{Scalable Video Coding}
\newacronym{tb}{TB}{Transport Block}
\newacronym{tcp}{TCP}{Transmission Control Protocol}
\newacronym{tdd}{TDD}{Time Division Duplexing}
\newacronym{tdm}{TDM}{Time Division Multiplexing}
\newacronym{tdma}{TDMA}{Time Division Multiple Access}
\newacronym{tfl}{TfL}{Transport for London}
\newacronym{tfrc}{TFRC}{TCP-Friendly Rate Control}
\newacronym{tft}{TFT}{Traffic Flow Template}
\newacronym{tgen}{TGEN}{Traffic Generator}
\newacronym{tip}{TIP}{Telecom Infra Project}
\newacronym{tm}{TM}{Transparent Mode}
\newacronym{to}{TO}{Telco Operator}
\newacronym{tr}{TR}{Technical Report}
\newacronym{trp}{TRP}{Transmitter Receiver Pair}
\newacronym{ts}{TS}{Technical Specification}
\newacronym{tti}{TTI}{Transmission Time Interval}
\newacronym{ttt}{TTT}{Time-to-Trigger}
\newacronym{tx}{TX}{Transmitter}
\newacronym{uas}{UAS}{Unmanned Aerial System}
\newacronym{uav}{UAV}{Unmanned Aerial Vehicle}
\newacronym{udm}{UDM}{Unified Data Management}
\newacronym{udp}{UDP}{User Datagram Protocol}
\newacronym{udr}{UDR}{Unified Data Repository}
\newacronym{ue}{UE}{User Equipment}
\newacronym{uhd}{UHD}{\gls{usrp} Hardware Driver}
\newacronym{ul}{UL}{Uplink}
\newacronym{um}{UM}{Unacknowledged Mode}
\newacronym{uml}{UML}{Unified Modeling Language}
\newacronym{upa}{UPA}{Uniform Planar Array}
\newacronym{upf}{UPF}{User Plane Function}
\newacronym{urllc}{URLLC}{Ultra Reliable and Low Latency Communications}
\newacronym{usa}{U.S.}{United States}
\newacronym{usim}{USIM}{Universal Subscriber Identity Module}
\newacronym{usrp}{USRP}{Universal Software Radio Peripheral}
\newacronym{utc}{UTC}{Urban Traffic Control}
\newacronym{vim}{VIM}{Virtualization Infrastructure Manager}
\newacronym{vm}{VM}{Virtual Machine}
\newacronym{vnf}{VNF}{Virtual Network Function}
\newacronym{volte}{VoLTE}{Voice over \gls{lte}}
\newacronym{voltha}{VOLTHA}{Virtual OLT HArdware Abstraction}
\newacronym{vr}{VR}{Virtual Reality}
\newacronym{vran}{vRAN}{Virtualized \gls{ran}}
\newacronym{vss}{VSS}{Video Streaming Server}
\newacronym{wbf}{WBF}{Wired Bias Function}
\newacronym{wf}{WF}{Waterfilling}
\newacronym{wg}{WG}{Working Group}
\newacronym{wlan}{WLAN}{Wireless Local Area Network}
\newacronym{osm}{OSM}{Open Source Management and Orchestration}
\newacronym{pnf}{PNF}{Physical Network Function}
\newacronym{drl}{DRL}{Deep Reinforcement Learning}
\newacronym{mtc}{MTC}{Machine-type Communications}
\newacronym{osc}{OSC}{O-RAN Software Community}
\newacronym{mns}{MnS}{Management Services}
\newacronym{ves}{VES}{\gls{vnf} Event Stream}
\newacronym{ei}{EI}{Enrichment Information}
\newacronym{fh}{FH}{Fronthaul}
\newacronym{fft}{FFT}{Fast Fourier Transform}
\newacronym{laa}{LAA}{Licensed-Assisted Access}
\newacronym{plfs}{PLFS}{Physical Layer Frequency Signals}
\newacronym{ptp}{PTP}{Precision Time Protocol}
\newacronym{asic}{ASIC}{Application-specific Integrated Circuit}
\newacronym{aal}{AAL}{Acceleration Abstraction Layer}
\newacronym{fec}{FEC}{Forward Error Correction}
\newacronym{sdl}{SDL}{Shared Data Layer}
\newacronym{nib}{NIB}{Network Information Base}
\newacronym{rnib}{R-NIB}{RAN \gls{nib}}
\newacronym{fcaps}{FCAPS}{Fault, Configuration, Accounting, Performance, Security}
\newacronym{ie}{IE}{Information Element}
\newacronym{fg}{FG}{Focus Group}
\newacronym{osfg}{OSFG}{Open Source Focus Group}
\newacronym{sdfg}{SDFG}{Standard Development Focus Group}
\newacronym{tifg}{TIFG}{Test \& Integration Focus Group}
\newacronym{sfg}{SFG}{Security Focus Group}
\newacronym{swg}{SWG}{Security Work Group}
\newacronym{e2sm}{E2SM}{E2 Service Model}
\newacronym{tsc}{TSC}{Technical Steering Committee}
\newacronym{sdo}{SDO}{Standard-Development Organization}
\newacronym{sql}{SQL}{Structured Query Language}
\newacronym{ssh}{SSH}{Secure Shell}
\newacronym{tls}{TLS}{Transport Layer Security}
\newacronym{netconf}{NETCONF}{Network Configuration Protocol}
\newacronym{dtls}{DTLS}{Datagram Transport Layer Security}
\newacronym{cmp}{CMP}{Certificate Management Protocol}
\newacronym{ccc}{CCC}{Cell Configuration and Control}
\newacronym{dsp}{DSP}{Digital Signal Processing}
\newacronym{opex}{OPEX}{Operational Expenses}
\newacronym{cbrs}{CBRS}{Citizen Broadband Radio Service}
\newacronym{ntn}{NTN}{Non-terrestrial Network}
\newacronym{gbr}{GBR}{Guaranteed Bitrate}
\newacronym{sps}{SPS}{Semi-Persistent Scheduling}
\newacronym{tbs}{TBS}{Transport Block Size}

\IEEEoverridecommandlockouts

\begin{document}
\bstctlcite{BSTcontrol}  

%
\title{Near-real-time and Dynamic Service Level Agreement Management with Open RAN}

\title{A Framework for Near-real-time and Dynamic Service Level Agreement Control with Open RAN}

\title{A Framework for Design and Evaluation of Near-real-time Open RAN Control}

\title{A Framework for Design and Evaluation of Dynamic Open RAN Control of 5G Base Station}

\title{A Framework for Design and Evaluation of Dynamic 5G Control with Open RAN}

\title{An Open RAN Framework for the Design and Evaluation of Dynamic 5G Control}

\title{An Open RAN Framework for Near-Real-Time\\5G Service Level Agreement Control}

\title{An Open RAN Framework for the Dynamic Control of 5G Service Level Agreements}



\author{
    \IEEEauthorblockN{Eugenio Moro\IEEEauthorrefmark{1}, Michele Polese\IEEEauthorrefmark{3}, Antonio Capone\IEEEauthorrefmark{1}, Tommaso Melodia\IEEEauthorrefmark{3}}\\
    \IEEEauthorblockA{\IEEEauthorrefmark{1}Department of Electronics, Information and Bioengineering, Polytechnic University of Milan, Italy\\
    \IEEEauthorrefmark{3}Institute for the Wireless Internet of Things, Northeastern University, Boston, MA, U.S.A\\
    \IEEEauthorrefmark{1}\{eugenio.moro, antonio.capone\}@polimi.it, \IEEEauthorrefmark{3}\{m.polese, t.melodia\}@northeastern.edu}
\thanks{This paper was partially supported by the NSF under Grant CNS-2117814.}
}


%


\maketitle

\begin{abstract}

The heterogeneity of use cases that next-generation wireless systems need to support calls for flexible and programmable networks that can autonomously adapt to the application requirements. Specifically, traffic flows that support critical applications (e.g., vehicular control or safety communications) often come with a requirement in terms of guaranteed performance. At the same time, others are more elastic and can adapt to the resources made available by the network (e.g., video streaming). 
To this end, the Open \gls{ran} paradigm is seen as an enabler of dynamic control and adaptation of the protocol stack of \gls{3gpp} networks in the \gls{5g} and beyond. Through its embodiment in the O-RAN Alliance specifications, it introduces the \glspl{ric}, which enable closed-loop control, leveraging a rich set of \gls{ran} \glspl{kpm} to build a representation of the network and enforcing dynamic control through the configuration of \gls{3gpp}-defined stack parameters. In this paper, we leverage the Open RAN closed-loop control capabilities to design, implement, and evaluate multiple data-driven and dynamic \gls{sla} enforcement policies, capable of adapting the \gls{ran} semi-persistent scheduling patterns to match users' requirements. To do so, we implement semi-persistent scheduling capabilities in the \gls{oai} \gls{5g} stack, as well as an easily extensible and customizable version of the Open \gls{ran} E2 interface that connects the \gls{oai} base stations to the near-real-time \gls{ric}. We deploy and test our framework on Colosseum, a large-scale hardware-in-the-loop channel emulator. Results confirm the effectiveness of the proposed Open RAN-based solution in managing \gls{sla} in near-real-time. 
\end{abstract}


%
\IEEEpeerreviewmaketitle

\glsresetall

\glsunset{nr} 

\section{Introduction}
\label{sec:intro}

As wireless networks improve toward ultra-high data rates, low latency, and high reliability, they also become essential to countless applications and use cases in our digital society. In a trend that started with the \gls{5g} of mobile networks and is continuing toward the \gls{6g}, the same air interface is used to serve extremely heterogeneous traffic patterns, with dynamic constraints and traffic loads~\cite{navarro2020survey}. As an example, cellular \glspl{ran} are designed to support \gls{embb} applications with an over-the-air frame structure, waveform, and protocol stack tailored to long-running data-rate-hungry streams (e.g., video streaming), but also more bursty traffic with low latency constraints or \gls{mtc} communications, thanks to configurations of the frame structure that favor short over-the-air bursts~\cite{dahlman20205g}. Similarly, elastic applications can use any amount of resources made available by the system, while other traffic flows require tight \gls{sla} and performance guarantees.

However, while the capabilities to support different traffic requirements are part of the technical specifications for 3GPP NR, the \gls{ran} for \gls{5g} systems, the actual commercial implementations lack the capability to dynamically and optimally switch between such configurations and to adapt to the actual user patterns and demand on the fly. This does not allow for efficient exploitation of the scarce spectrum resources available to wireless networks, and leads to a mismatch between user expectations and achievable performance~\cite{zhang2023dependent}.
A typical example is represented by the significant body of literature on how to optimally select waveform parameters~\cite{patriciello2019e2e,lien20175g} and enforce \gls{sla} constraints~\cite{liu2021onslicing,chien2020end,abouaomar2023federated} in cellular systems, which however is often not deployable in real scenarios and \gls{ran} stacks due to their inflexibility and limited adaptability. 

The recent paradigm shift introduced by the Open \gls{ran} vision is implementing practical primitives for the dynamic adaptation and optimization of the \gls{ran} configurations, enabling the adoption of more flexible solutions for the support of different traffic requirements~\cite{bonati2021intelligence}. Specifically, Open \gls{ran} introduces new components, the \glspl{ric}, which interact with 3GPP-compliant base stations through open interfaces, and have the capability to (i) receive telemetry and \glspl{kpm} from the \gls{ran}; (ii) infer the status of the system using data-driven approaches; and (iii) apply new configurations to the radio resource management process and adapt the \gls{ran} behavior to the actual conditions on the ground~\cite{polese2023understanding}. The Open \gls{ran} vision, and its embodiment in the O-RAN Alliance specifications, include two \glspl{ric} for near-real-time (10 ms---1 s) and non-real-time (more than 1 s) closed-loop control, implemented through xApps and rApps, respectively.

In this paper, we build on the Open \gls{ran} vision and identify and compare two control strategies that an operator can use to enforce \gls{sla} for non-elastic traffic, to be deployed as custom control logic in the near-real-time \gls{ric}. The strategies build on 3GPP- and O-RAN-compliant parameters that allow specifying semi-persistent scheduling patterns in 5G base stations, and thus they do not require modifications in protocol stacks that comply with technical specifications. Specifically, the two \gls{sla} management solutions embody either a \textit{strict} or a \textit{soft} \gls{sla} enforcement policy, which can be selected according to the network operator objectives.

The second contribution of this paper is an implementation of the experimental infrastructure required to prototype, test, and evaluate such control logic in an end-to-end, programmable framework for the design of custom Open \gls{ran} closed-loop control. Specifically, we extend the OpenRAN Gym platform, first introduced in~\cite{bonati2022openrangym_pawr}, to connect the open-source OpenAirInterface (OAI) 5G \gls{ran} implementation~\cite{kaltenberger2020openairinterface} to a near-real-time \gls{ric} based on the O-RAN Software Community distribution~\cite{oran-wg3-ricarch}. This combines a state-of-the-art Open RAN platform with the feature-rich and standard-compliant 5G OAI implementation. Our implementation of the E2 interface connecting the RIC to the RAN is designed to be easily extensible, thanks to an abstraction of its functionalities based on human-readable data structures that can be automatically compiled into serializable buffers. We also extend the OAI stack to support the semi-persistent scheduling control and integrate it with our E2 implementation. 

Finally, we profile the performance of the combined infrastructure and control logic in Colosseum, the world's largest wireless network emulator with hardware in the loop. We show how the entire framework, comprising the RAN nodes, the near-RT RIC and the xApps, is highly effective at controlling the SLA enforcement at the timescale of 100 ms.

The rest of the paper is organized as follows. Section~\ref{sec:oran} presents an overview of the \gls{oran} paradigm and its associated architecture. Section~\ref{sec:control} details the modifications to \gls{oai}, the xApp \gls{sdk} and their integration in OpenRAN Gym as a complete experimental framework. Section~\ref{sec:sla_policies} details the two SLA policies and their implementation as \gls{oran} control micro-services. Finally, Section~\ref{sec:eval} presents the numerical evaluation of the entire framework on a large-scale channel emulator with hardware in the loop. 

\section{Open \gls{ran} - A Primer}
\label{sec:oran}

\begin{figure}[t]
    \centering
    \includegraphics[width=\columnwidth]{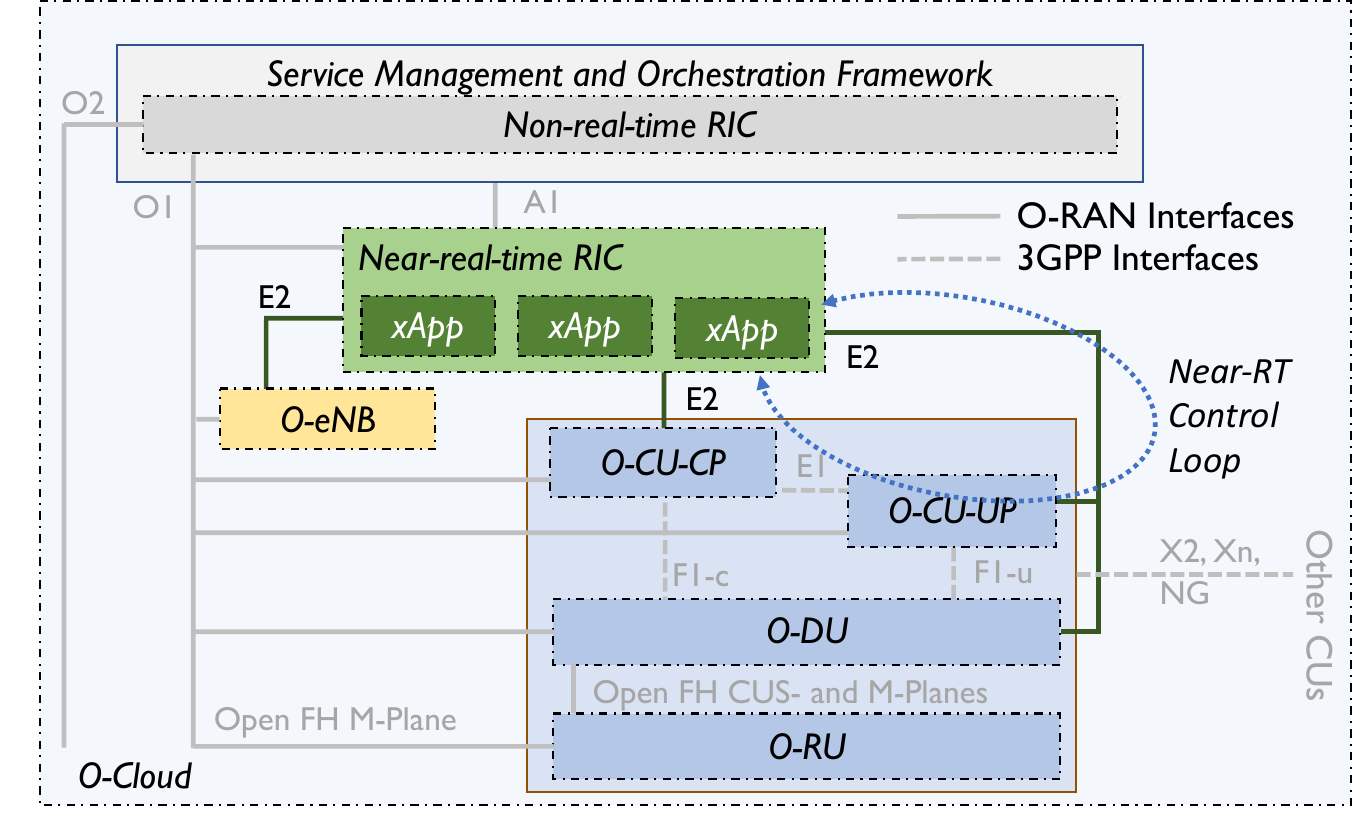}
    \caption{Logical view of the O-RAN architecture, adapted from~\cite{polese2023understanding}. The focus is on the E2 interface, connecting the near-RT \gls{ric} and the \gls{ran} nodes for the near-RT control applications discussed in this paper.}
    \label{fig:o-ran-architecture}
\end{figure}


The Open \gls{ran} paradigm is implemented by the O-RAN Alliance, an industry and academic consortium with more than 300 members, in the architecture shown in Fig.~\ref{fig:o-ran-architecture}. In this, the \gls{ran} \glspl{gnb} are disaggregated and split into multiple nodes with different functionalities based on the layers of the protocol stack they host. The \gls{sdap}, \gls{pdcp}, and \gls{rrc}, i.e., the higher layers for the \gls{up} and the \gls{cp}, are in the \gls{cu}-UP and CU-CP, respectively. The \gls{du} hosts three layers that operate in a tightly synchronous fashion, the \gls{rlc}, the \gls{mac}, and the higher part of the physical layer. Finally, the \gls{ru} features the lower part of the physical layer and the \gls{rf} frontend. These nodes are connected to each other using 3GPP-defined interfaces and the Open Fronthaul interface from the O-RAN Alliance.

In addition, they are connected through the E2 and the O1 interfaces to the near-RT and non-RT \glspl{ric}, respectively. These intelligent controllers can onboard plug-and-play control logic (i.e., xApps and rApps) to extend the functionalities of \gls{ran} nodes with custom control loops. Specifically, the non-RT RIC, embedded in the \gls{smo}, relies on rApps to perform policy and control updates in non-real time, i.e., with a time granularity higher than 1 s. This is to dynamically update configurations in the \gls{ran} nodes and control high-level policies and parameters, e.g., cell sleeping patterns in the CU/DU and beam codebooks at the RU. The near-RT RIC, instead, uses xApps to implement control loops that execute in less than 1 s, but more than 10 ms. The near-RT RIC closes the control loop to perform radio resource management in the DU and CU. Consequently, it operates on parameters that need to be updated with more stringent deadlines compared to the non-RT control loop. 

The near-RT RIC relies on the E2 interface: a logical point-to-point interface running on top of SCTP. The E2 interface features two components. The first is an application protocol, or E2AP, which manages the connection between each DU/CU (also known as E2 nodes) and the near-RT RIC, including setup, monitoring, and teardown. The E2AP offers a set of primitives (e.g., indication, report, and control messages) assembled to build custom E2 \emph{service models}, or E2SMs. The E2SM is a component implemented on top of E2AP which provides the semantics to the E2 interface, e.g., reporting of \glspl{kpm} from the \gls{ran} with E2SM \gls{kpm}, or control of \gls{ran} parameters with E2SM \gls{rc}. Specifically, \gls{rc} has been designed to interact with protocol stack parameters defined in 3GPP specifications, introducing an elegant solution to decouple control from the actual \gls{ran} implementation. 

\section{A Framework for 5G Open \gls{ran} Closed-Loop Control}
\label{sec:control}

This section discusses how we implemented an easily extensible E2 interface for the \gls{oai} 5G protocol stack and a companion xApp \gls{sdk} which leverages O-RAN-compliant E2AP and custom E2SMs. Both are open-source and publicly available,\footnote{Refer to the \url{openrangym.com} website for links to code and tutorials.} and enable design and testing the dynamic \gls{sla} policies we propose and evaluate in this paper.

\subsection{Integrating OpenRAN Gym and OpenAirInterface}
\label{sec:oai}

\gls{oai} is an open-source project that provides the implementation of a \gls{3gpp} Release 15 NR \gls{ran} and 5G Core~\cite{kaltenberger2020openairinterface}. The \gls{ran} base stations can be deployed on generic compute hardware and leverage software-defined radios via multiple drivers, or O-RAN \glspl{ru} with a 7.2x fronthaul implementation. Being open, programmable, and easily extensible, \gls{oai} can be leveraged to implement and test O-RAN closed-loop control in a 5G-compliant end-to-end environment. 

OpenRAN Gym represents the first publicly accessible platform that features \gls{oran}-based data collection and control frameworks for data-driven experimentation at scale~\cite{bonati2022openrangym_pawr}. 
In particular, OpenRAN Gym frameworks package the entire software chain required to deploy the \gls{oran} components presented in Fig.~\ref{fig:o-ran-architecture}, namely the base stations, the near-RT RIC, and the xApps. The RIC is based on the \gls{osc} RIC implementation, which has been adapted for deployment on a variety of open testbeds, including Colosseum (Sec.~\ref{sec:eval}) and the \gls{pawr} testbeds COSMOS and POWDER.

\begin{figure}
    \centering
    \includegraphics[width=\columnwidth]{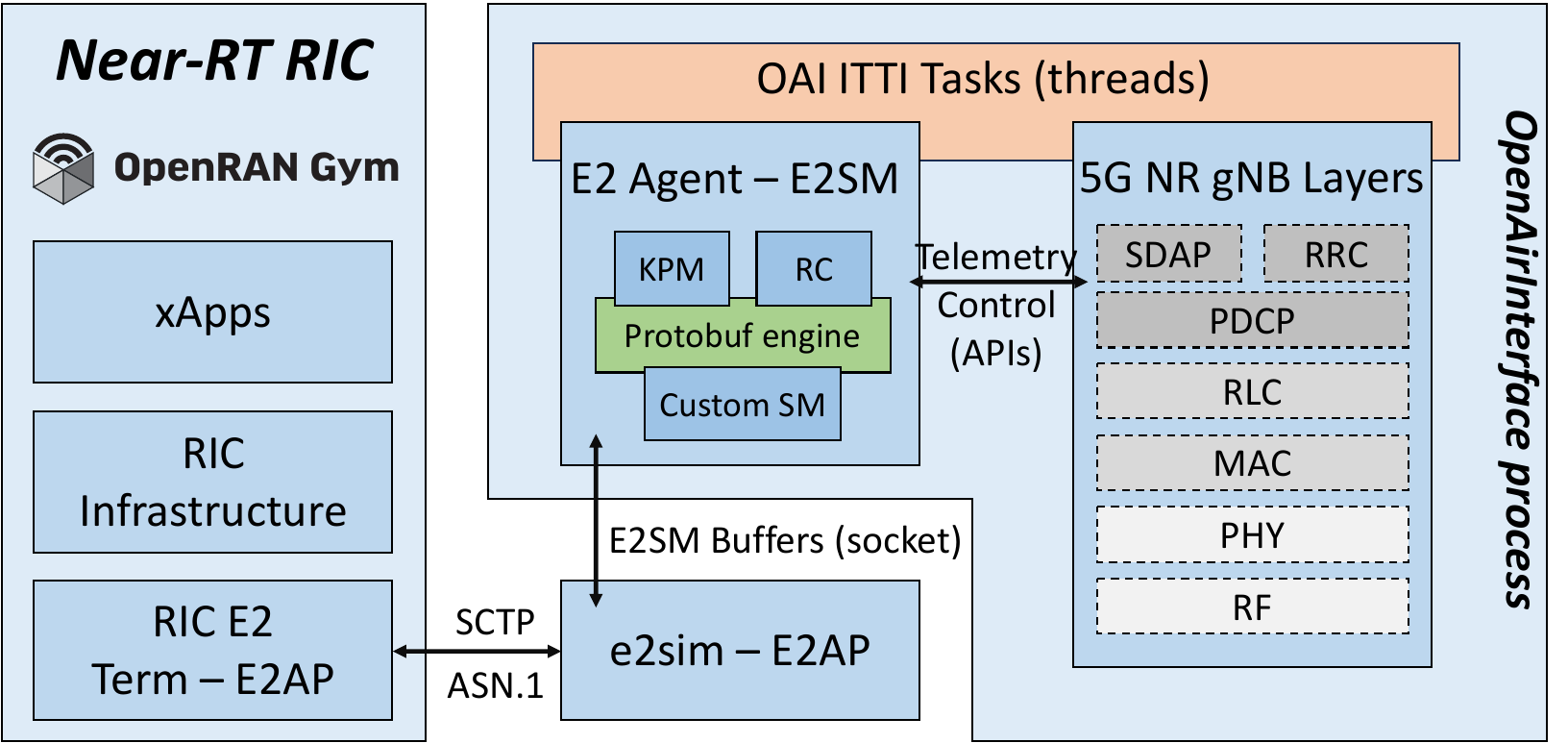}
    \caption{Integration of the E2 interface in OAI.}
    \label{fig:e2-oai}
\end{figure}

Figure~\ref{fig:e2-oai} illustrates how we designed and implemented the integration between \gls{oai} and the OpenRAN Gym framework through the E2 interface, enabling experiments with a \gls{3gpp}-compliant, \gls{oran}-enabled 5G \gls{sa} deployment. 
As discussed in Sec.~\ref{sec:oran}, the E2 interface is functionally split into 2 sub-protocols: E2AP and E2SM. Similarly, our custom E2 agent has been split into 2 components, according to the principle of separation of responsibilities. The E2AP component runs as a standalone application, and it is based on E2AP libraries extracted from \gls{osc} E2 simulator library (\texttt{e2sim})~\cite{e2sim_webpage}. This component manages the E2AP connection lifecycle with the near-RT RIC. It is standard-compliant, i.e., it encodes and decodes E2AP messages based on O-RAN ASN.1 definitions and an SCTP transport layer.

The component that provides E2SM functionalities is integrated into \gls{oai} codebase and runs as a task inside the \gls{gnb} process, similarly to the implementation in~\cite{chen2023flexapp,schmidt2021flexric}. Consequently, the E2 agent implementing the E2SM has direct access to the \gls{gnb} data structures and processes and can effectively perform data collection and apply control actions by directly interacting with the variables that define the relevant 3GPP parameters to tune.

The two components run as independent processes in the same machine (potentially providing a resilient E2 interface in case of failures in the \gls{gnb}), and they communicate through UDP sockets. When the E2AP component receives an E2AP message from the near-RT RIC, the E2SM payload is extracted, decoded according to the ASN.1 schema, and sent to the OAI E2 agent. Here the E2SM payload is further decoded and processed. Similarly, any E2SM payload produced by the E2SM component in the \gls{gnb} is sent to \texttt{e2sim}, which handles E2AP encapsulation and message delivery.  

In this architecture, the E2SM payload is a buffer of bytes without additional specifications or requirements. Therefore, it can be an O-RAN-compliant E2SM ASN.1-encoded payload, a custom unstructured string, or---as we propose in this paper---a protobuf buffer. Protobuf, or Protocol Buffers, is a data serialization format developed by Google to efficiently exchange structured data between different systems~\cite{currier2022protocol}. It uses a language-agnostic schema to define the data structures, which can then be compiled into specific implementations in different programming languages (e.g., C for the OAI E2 agent and Python for the xApp described in Sec.~\ref{sec:xapp}). Compared to ASN.1, protobuf data structures are more user-friendly to extend, compile, and integrate into the code, making it a practical tool for developing custom E2 service models for research and testing. 

\subsection{A Python Framework for Flexible RAN Control xApps}
\label{sec:xapp}

\begin{figure}[t]
    \centering
    \includegraphics[width=.9\columnwidth]{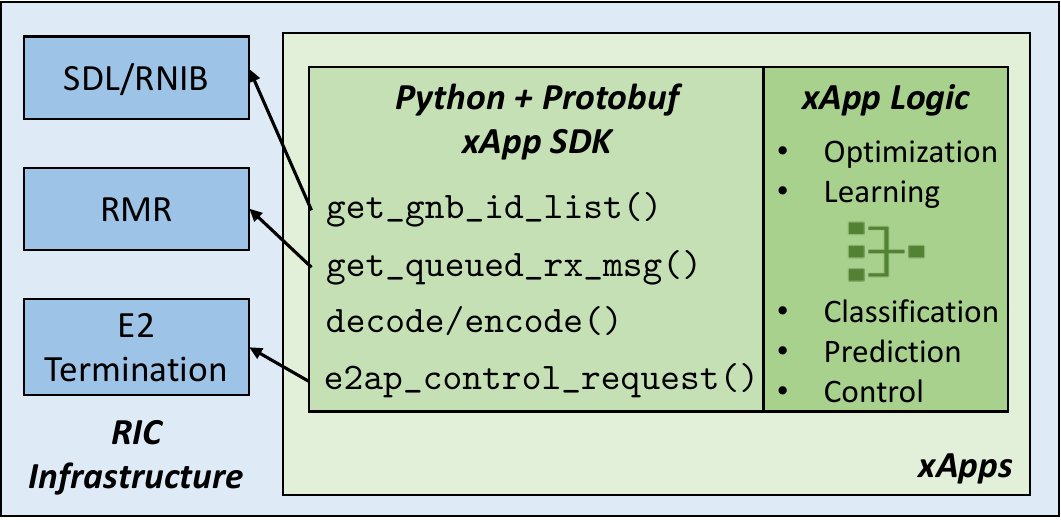}
    \caption{Python framework for flexible xApp development.}
    \label{fig:xapp}
\end{figure}

To streamline and simplify the xApp development process, we developed and made publicly available\footnote{\url{https://github.com/ANTLab-polimi/xapp-e2ap-py}} an xApp \gls{sdk} in Python, a popular programming language which is also used in several state-of-the-art frameworks for data-driven inference~\cite{tensorflow2015-whitepaper}. Figure~\ref{fig:xapp} illustrates at a high level the structure of the xApp and its \gls{sdk}, where the xApp logic leverages the \gls{sdk} for operations related to data encoding/decoding and interaction with the rest of the \gls{ric} platform.

We use the \gls{osc} Python xApp framework as a library to wrap primitives for the communication with the other internal \gls{ric} components, i.e., the E2 termination to the \gls{ran} nodes, the \gls{rmr}, which dispatches internal messages across xApps and the RIC infrastructure, and the \gls{sdl}, containing multiple data repositories with information on the E2 nodes (e.g., the \gls{rnib}). The xApp \gls{sdk} wraps the code required to interact with these components into functions that the developer can easily embed in their xApp logic. Some examples are shown in Fig.~\ref{fig:xapp}. The \texttt{get\_gnb\_id\_list()} \gls{api} retrieves the list of \gls{ran} nodes connected over E2, which can be either leveraged to retrieve \glspl{kpm} or as targets for the control and optimization; \texttt{get\_queued\_rx\_msg()} is a non-blocking method that parses the queue of incoming E2 messages for the xApp; and \texttt{e2ap\_control\_request()} sends a message to the E2 termination with a control request in its buffer. 

In addition, as discussed in Sec.~\ref{sec:oai}, the xApp natively embeds the \glspl{api} to create, encode, and decode messages following the protobuf format. The same definition can be added to the xApp and the E2 agent in the \gls{ran} and properly compiled in the desired programming language. 

For this paper, we implement a service model equivalent to E2SM RC to control and optimize the semi-persistent scheduling patterns in the \gls{gnb}. Further details are provided in Sec.~\ref{sec:sla_policies}, where we describe the optimization policies and their enforcement through the RC-based service model.

\section{Open \gls{ran} Control --- \gls{sla} Enforcement Policies and Control}
\label{sec:sla_policies}

This section describes the two \gls{sla} enforcement strategies we implement on top of the Open RAN framework described in Sec.~\ref{sec:control}, leveraging the implementation of an E2SM based on \gls{rc} for \gls{gbr} control in \gls{oai} (Sec.~\ref{sec:rcoai}) and a data-driven optimization framework implemented the \gls{sla} xApp (Sec.~\ref{sec:strategies}).


\subsection{\gls{rc} for GBR in OAI.}
\label{sec:rcoai}

The \gls{sla} xApp leverages data collection and control primitives to provide a guaranteed bitrate to the \glspl{ue}.
The xApp needs information on the cell resource utilization and the per-UE channel quality, and it needs to inform the \gls{gnb} \gls{mac} layer about how many resources should be allocated to each \gls{ue}. This is achieved through E2 service models.

The \gls{oran} \gls{rc} and \gls{kpm} service models provide primitives to query telemetry from the \gls{gnb}. Specifically, \gls{rc} provides the \gls{tbs} information, namely the bits that are exchanged between the \gls{gnb} and each \gls{ue} in a \gls{tti}. This information can be used in the xApp to estimate the \gls{ue} MAC throughput with extreme precision.
In \gls{5g} \gls{nr}, the \gls{gnb} resources are partitioned in \glspl{prb}, i.e., the minimum unit of spectrum and time resources that can be allocated to \glspl{ue} at each \gls{tti}. Baseides the \gls{tbs} information in \gls{rc}, \gls{kpm} exposes the \gls{prb} allocation information at the \gls{ue} level, which the xApp requires to know the \gls{gnb} resource distribution. 

By combining the \gls{mac} throughput and \gls{prb} allocation information, the xApp can compute the \textit{per-\gls{prb}} throughput for each \gls{ue}. This measure is a proxy for the \gls{ue} spectral efficiency, and it can be used to inform the allocation decisions behind any \gls{sla} management policy.  

Once the \gls{gbr} allocation decision is taken in the xApp (with the policies discussed in Sec.~\ref{sec:strategies}), it must be enforced in the \gls{gnb}. To this end, we leverage the E2SM \gls{rc} capabilities of controlling the \gls{sps} process. In \gls{5g} \gls{nr}, \gls{sps} is a mechanism that allows the \gls{gnb} to schedule parts of the resources on a fixed and persistent basis, as opposed to the more traditional dynamic scheduler where resources are granted based on traffic conditions.
By properly configuring \gls{sps}, the xApp can reserve the fixed resource portion required by each \gls{ue} to obtain the desired \gls{gbr}. Note that this is an NR-compliant feature, thus, by controlling this with E2SM \gls{rc}, it is possible to practically implement dynamic \gls{gbr} policies for \gls{sla} enforcement in 5G \glspl{gnb}.

The \gls{oai} 5G \gls{mac} implementation does not support \gls{sps}, neither it supports the necessary \glspl{api} to control it through \gls{rc}. As previously mentioned, we have implemented a lightweight \gls{rc} service model that exposes the required data collection and control knobs into \gls{oai}. In particular, the \gls{prb} allocation and \gls{tbs} information are collected inside \gls{oai} MAC implementation by adding the required hooks in both the downlink and uplink schedulers. Additionally, we provide \gls{sps} support in \gls{oai} by modifying the aforementioned schedulers so that the \gls{tbs} of any \gls{ue} can be fixed through our custom \gls{rc} control messages, at any time. This effectively results in a fixed per-\gls{ue} resource allocation, as one would have with \gls{sps}. 

\subsection{Dynamic \gls{sla} Enforcement Strategies}
\label{sec:strategies}

We leverage the dynamic closed-loop control framework provided by O-RAN to design, implement, and evaluate two \gls{sla} enforcement strategies that dynamically manage radio resources with the goal of providing a guaranteed bitrate to \glspl{ue} and data flows that require it. 
Through the data collection capabilities of the custom \gls{rc} service model, the xApp can enforce allocations for a \gls{gbr}, as long as enough resources are available. Whenever this is not true, the system cannot support all the \gls{gbr} requests, and the resources must be managed according to specific resource contention resolution policies to minimize the \gls{sla} violation. For a specific \gls{ue} $u$, we define its \gls{sla} violation as the difference $v_u$ between the throughput $\text{SLA}_u$ requested as the guaranteed \gls{gbr} value and the experienced throughput $p_u \eta_u$, with $p_u$ amount of allocated \glspl{prb} and $\eta_u$ the per-PRB throughput.

We propose and evaluate two policies. The first is a flexible \gls{sla} enforcement (policy \emph{Soft}). When resource contention resolution is required, the xApp dynamically allocates the available \glspl{prb} such that the overall sum of the per-\gls{ue} \gls{sla} violation $v_u$ is minimized. This policy can be expressed through the following linear program: 
\begin{align}
    &\min \sum_{u\in \mathcal{U}} v_u,&\label{opt:elast:p1}\\
    &\sum_{u \in \mathcal{U}} p_u \leq C,&\label{opt:elast:cap}\\
    & v_u \geq \text{SLA}_u - p_u \eta_u&\forall u \in \mathcal{U},\label{opt:elast:viol}\\
    &v_u, p_u \geq 0&\forall u \in \mathcal{U}.
\end{align}
Here $\mathcal{U}$ represents the set of \gls{gbr} \glspl{ue} associated with the \gls{gnb}, and $v_u, \text{SLA}_u, p_u, \eta_u$ are defined as above.
Constraint~\eqref{opt:elast:cap} limits the overall allocated \glspl{prb} to the maximum amount $C$ of \glspl{prb} available to the \gls{gnb}.

The second policy is based on a more rigid \gls{gbr} enforcement (policy \emph{Strict}). According to this policy, each \gls{ue} $u$ is assigned with a \emph{weight} $w_u$ representing priority, economic value, or other importance metrics. In case of resource contention, the policy selects a subset of \glspl{ue} to continue to serve with \gls{gbr} such that the overall weight (i.e., the sum of selected \glspl{ue} weights) is maximized. The \glspl{ue} outside the optimal subset are either given the remaining resources or disconnected from the \gls{gnb}. This policy can be mathematically expressed through a knapsack formulation~\cite{salkin1975knapsack} using the previous notation, as follows:
\begin{align}
    &\max \sum_{u \in \mathcal{U}} x_u w_u,&\\
    &\sum_{u \in \mathcal{U}}x_u c_u \leq C,\label{opt:knap:cap}\\
    &x_u \in \{0,1\}&\forall u \in \mathcal{U}.
\end{align}
Here $x_u$ is a binary variable indicating whether the \gls{ue} is inside the \gls{gbr}-enforced subset or not. Parameter $c_u$ represents the \gls{gnb} resources required to guarantee the \gls{sla} for \gls{ue} $u$, considered in the capacity constraint~\eqref{opt:knap:cap}.

\section{Experimental Evaluation}
\label{sec:eval}

This section describes the experimental setup and evaluation of the dynamic \gls{sla} policies on Colosseum and \gls{oai}.

\subsection{The Colosseum Testbed}

We evaluated the solutions discussed in Sec.~\ref{sec:sla_policies}---together with the xApp-based control framework for OAI---on Colosseum, the world's largest wireless network emulator with hardware-in-the-loop~\cite{bonati2021colosseum}. The Colosseum testbed provides users with access to 128 pairs of servers and USRP X310 (collectively defined as \glspl{srn}). The server can load an LXC container with a custom image provided by the user (e.g., the OpenRAN Gym near-RT RIC, the OAI gNB, and OAI UE). The radio is connected to Colosseum's \gls{mchem}, which implements virtual \gls{rf} scenarios with path loss, fading, and interference by filtering the transmitted signal from the \gls{srn} radio with the channel impulse response. \gls{mchem} leverages a bank of 64 \glspl{fpga} and can emulate channels with up to 4 multi path components and pre-defined mobility for the nodes.

Colosseum has already been widely used to evaluate custom control logic for Open RAN systems through OpenRAN Gym. Here, we deploy a network with 6 nodes, including a \gls{5g} \gls{cn} node, one \gls{gnb}, 3 \glspl{ue}, and a near-RT RIC. The \gls{rf} scenario emulates a typical laboratory testing environment, with a fixed pathloss among the \gls{ran} nodes. The \gls{gnb} is configured for transmitting at 3.6 GHz in band n78, with a bandwidth of 40 MHz and numerology 1. \textit{iperf} generates full-buffer downlink TCP traffic. We limit our analysis to the downlink only case. The xApp is configured to adjust the resource allocation every 100 ms. 

\subsection{Results}

We configure the three \glspl{ue} with different \gls{gbr} values, i.e., 15 Mbps for UE1, 10 Mbps for UE2, and 5 Mbps for UE3. The \gls{gnb} can allocate 65 \glspl{prb}, which are not sufficient to satisfy \gls{sla} for all the \glspl{ue}.
We sequentially generate full-buffer traffic (starting from UE1, to UE2, and then UE3) and analyze the system performance. During the experiments, the xApp actively enforces a policy described in Sec.~\ref{sec:sla_policies}. Additionally, we have included a baseline in which the xApp is inactive and the \glspl{ue} allocations follow a proportional fairness scheduler. In every case, we measure the experienced throughput and the \gls{sla} violation, i.e., the difference between the nominal \gls{gbr} value and the actual throughput. 

\begin{figure*}
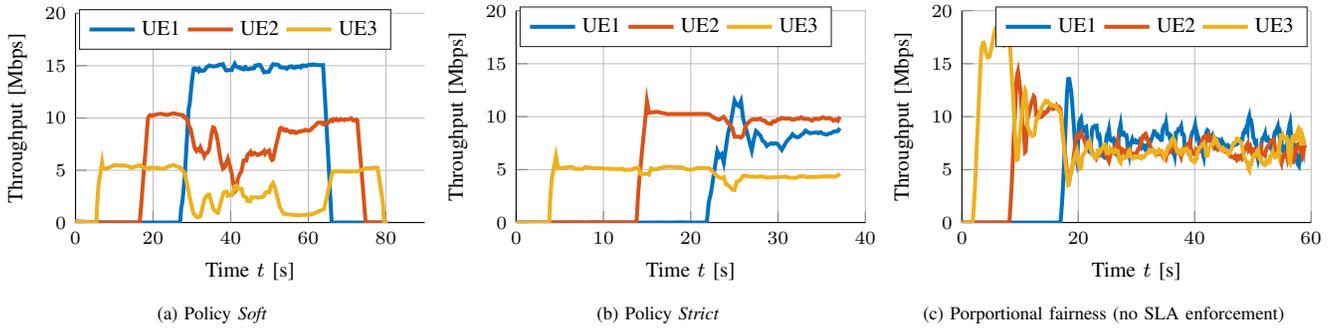

\begin{subfigure}[b]{.32\linewidth}
    \centering
    \setlength\fwidth{.8\columnwidth}
    \setlength\fheight{.5\columnwidth}
    \input{figures/elastic.tex}
    \caption{Policy \emph{Soft}}
    \label{fig:thr_p1}
\end{subfigure}
\begin{subfigure}[b]{.32\linewidth}
    \centering
    \setlength\fwidth{.8\columnwidth}
    \setlength\fheight{.5\columnwidth}
    \input{figures/knapsack.tex}
    \caption{Policy \emph{Strict}}
    \label{fig:thr_p2}
\end{subfigure}
\begin{subfigure}[b]{.32\linewidth} 
      \centering
    \setlength\fwidth{.8\columnwidth}
    \setlength\fheight{.5\columnwidth}
    \input{figures/prop_fairness.tex}
    \caption{Porportional fairness (no \gls{sla} enforcement)}
    \label{fig:thr_p3}
\end{subfigure}
\caption{Evolution of the per-UE throughput during different experiments, where traffic is started in sequence for UE3, UE2, and UE1.}
\label{fig:thr}
\end{figure*}

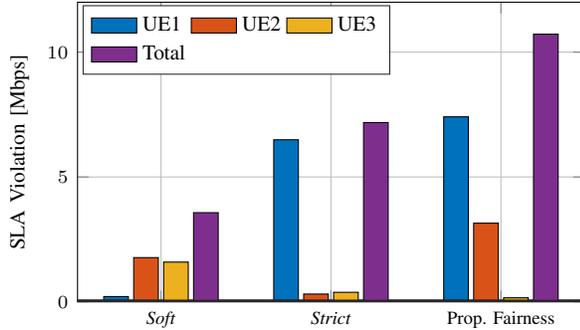
\begin{figure}[t]
    \centering
    \setlength\fwidth{.8\columnwidth}
    \setlength\fheight{.45\columnwidth}
%
%
\definecolor{mycolor1}{rgb}{0.00000,0.44700,0.74100}%
\definecolor{mycolor2}{rgb}{0.85000,0.32500,0.09800}%
\definecolor{mycolor3}{rgb}{0.92900,0.69400,0.12500}%
\definecolor{mycolor4}{rgb}{0.49400,0.18400,0.55600}%
\begin{tikzpicture}

\begin{axis}[%
width=0.951\fwidth,
height=\fheight,
at={(0\fwidth,0\fheight)},
scale only axis,
bar shift auto,
xmin=0.509090909090909,
xmax=3.49090909090909,
xtick={1,2,3},
xticklabels={{\emph{Soft}},{\emph{Strict}},{Prop. Fairness}},
ymin=0,
ymax=12,
ylabel style={font=\bfseries\color{white!15!black}},
ylabel={SLA Violation [Mbps]},
axis background/.style={fill=white},
ymajorgrids,
my axis style,
legend columns=3,
legend style={legend cell align=left, align=left, draw=white!15!black,font=\footnotesize, anchor=north west, at={(0.01,0.99)}},
]
\addplot[ybar, bar width=0.145, fill=mycolor1, draw=black, area legend] table[row sep=crcr] {%
1	0.21\\
2	6.49\\
3	7.41\\
};
\addplot[forget plot, color=white!15!black, line width=1.5pt] table[row sep=crcr] {%
0.509090909090909	0\\
3.49090909090909	0\\
};
\addlegendentry{UE1}

\addplot[ybar, bar width=0.145, fill=mycolor2, draw=black, area legend] table[row sep=crcr] {%
1	1.77\\
2	0.31\\
3	3.15\\
};
\addplot[forget plot, color=white!15!black, line width=1.5pt] table[row sep=crcr] {%
0.509090909090909	0\\
3.49090909090909	0\\
};
\addlegendentry{UE2}

\addplot[ybar, bar width=0.145, fill=mycolor3, draw=black, area legend] table[row sep=crcr] {%
1	1.59\\
2	0.38\\
3	0.16\\
};
\addplot[forget plot, color=white!15!black, line width=1.5pt] table[row sep=crcr] {%
0.509090909090909	0\\
3.49090909090909	0\\
};
\addlegendentry{UE3}

\addplot[ybar, bar width=0.145, fill=mycolor4, draw=black, area legend] table[row sep=crcr] {%
1	3.57\\
2	7.18\\
3	10.72\\
};
\addplot[forget plot, color=white!15!black, line width=1.5pt] table[row sep=crcr] {%
0.509090909090909	0\\
3.49090909090909	0\\
};
\addlegendentry{Total}

\end{axis}
\end{tikzpicture}%
    \caption{SLA violation, i.e., difference between requested \gls{gbr} and actual measured throughput, per \gls{ue} and total.}
    \label{fig:barplot}
\end{figure}

We start by analyzing the system behaviour when policy \emph{Soft} is active (Fig.~\ref{fig:thr_p1}). 
At the beginning of the experiment, only UE3 is receiving traffic and its \gls{sla} is met. UE2 becomes active at time $t=19$ s, but the resources are still sufficient to meet both \glspl{sla}. When UE1 becomes active, it requests resources for 15 Mbps, exceeding the available \glspl{prb} and activating the contention resolution mechanism of the xApp. In this case, the optimization formulation of policy \emph{Soft} selects UE1 to obtain the full \gls{gbr}, while UE2 and UE3 obtain less-than-required resources, but the overall throughput degradation is limited, following the goal of policy \emph{Soft} (minimize the overall \gls{sla} violation, Eq.~\eqref{opt:elast:p1}).
This is also confirmed by Fig.~\ref{fig:barplot}, which reports the per-UE and total \gls{sla} violation and shows how policy \emph{Soft} has a smaller overall violation with respect to policy \emph{Strict} and the baseline, at the expense of higher per-UE violation for UE2 and UE3.

Similarly, for policy \emph{Strict} (Fig.~\ref{fig:thr_p2}), the xApp contention resolution mechanism is required at time $t=21$ s, when UE1 starts exchanging traffic. In this case, however, policy \emph{Strict} results in UE2 and UE3 being served their full \gls{gbr}, while the remaining \gls{gnb} resources are left to UE1. This is expected since all the \glspl{ue} are set with the same weight and, thus, the xApp simply maximizes the number of satisfied \glspl{ue}. As shown in Fig.~\ref{fig:barplot}, for policy \emph{Strict}, most of the overall violation comes from UE3, while UE1 and UE2 have a negligible level of \gls{sla} violation.

Finally, we compare the two policies with the baseline, where no bitrate is guaranteed, and the resources are allocated according to proportional fairness scheduling (Fig.~\ref{fig:thr_p3}). In this case, all the \glspl{ue} obtain the same average throughput of around 7.5 Mbps. Consequently, UE1 and UE2 experience a high \gls{sla} violation, while UE3 does not, as its throughput is higher than the \gls{gbr} value for most of the duration of the experiment.

Overall, these experiments show how the proposed framework enables an effective implementation of near-RT dynamic \gls{sla} management mechanisms, which can be directly interfaced with real \gls{5g} \gls{ran} deployments. 

\section{Conclusions}
\label{sec:conclusions}
This paper presented experimental capabilities that enable data-driven \gls{oran} experimentation at scale with an open-source 5G implementation based on \gls{oai}. This has been packaged into the OpenRAN Gym framework, including an \gls{oran} compatible extension of \gls{oai}, with an easily extensible E2 agent, and an xApp SDK compatible with the \gls{osc} near-RT \gls{ric}. We have leveraged on this framework to build a near-real-time \gls{oran}-based dynamic \gls{gbr} \gls{sla} management solution, where two different resource contention resolution policies are implemented as xApps. We have deployed the framework on Colosseum, and results demonstrate the effectiveness of both the proposed \gls{sla} management solution and the framework as a whole. 

\footnotesize
\vspace{-0.5cm}
\bibliographystyle{IEEEtran}
\bibliography{biblio.bib} 

\end{document}